\documentclass[9pt,twocolumn,twoside]{opticajnl}

\journal{ol} 

\setboolean{shortarticle}{true}

\usepackage{lineno}

\title{Optimal optical Ferris wheel solitons in a nonlocal Rydberg medium}

\author[1]{Jia-Bin Qiu}
\author[2,*]{Lu Qin}
\author[2,$\dagger$]{Xing-Dong Zhao}
\author[1,$\ddagger$]{Jing Qian}

\affil[1]{State Key Laboratory of Precision Spectroscopy, Department of Physics, School of Physics and Electronic Science, East China Normal University, Shanghai 200241, China}
\affil[2]{School of Physics, Henan Normal University, Xinxiang 453000, China}

\affil[*]{Corresponding author: qinlu@htu.edu.cn}
\affil[$\dagger$]{Corresponding author: phyzhxd@gmail.com}
\affil[$\ddagger$]{Corresponding author: jqian1982@gmail.com}

\begin{abstract}
We propose a scheme for the creation of
stable optical Ferris wheel(OFW) solitons in a nonlocal Rydberg electromagnetically induced transparency(EIT) medium. Depending on a careful optimization to both the atomic density and the one-photon detuning, we obtain an appropriate 
nonlocal potential provided by the strong interatomic interaction in Rydberg states which can perfectly compensate for the diffraction of the probe OFW field. Numerical results show that the fidelity keeps larger than 0.96 while the propagation distance has exceeded 160 diffraction lengths. Higher-order OFW solitons with arbitrary winding numbers are also discussed.
Our study provides a straightforward route to generate spatial optical solitons in the nonlocal response region of cold Rydberg gases.
\end{abstract}

\setboolean{displaycopyright}{true}

\begin{document}

\maketitle



{\it Introduction-} Solitons are a special class of localized fields that can maintain their spatial profiles as they propagate. Realizing three-dimensional spatial optical solitons is a long-standing goal in the study of nonlinear optics \cite{RevModPhys.83.247}. However, when the optical field carries orbital angular momentum(OAM) {\it i.e.} a so-called vortex beam, it tends to split up during propagation
due to the presence of a phase singularity at the origin \cite{OE.20.022961,NJP.20-043023,PRL.129.073902}.
So higher-order optical vortex beams are inherently unstable unless other singularity compensation approaches are used \cite{OE.22.009920,OPTICA.3.000355}.
Optical Ferris wheel(OFW), is
formed by the superposition of two vortex beams with opposite winding numbers \cite{PRA.96.013622}, which has attracted considerable attention because of its spectacular contribution in the atomic trapping and coherent control \cite{OE.15.008619,JOSAB.443903,OL.427000}. Unlike a pure vortex beam, OFW exhibits cylindrical symmetry only in intensity which would not occur singularity splitting as it propagates \cite{nanoph-2021-0814}. To create stable OFW solitons, a major challenge is to seek for a suitable nonlinear medium with tunable diffraction and nonlocal nonlinearity. 
When the diffraction of the optical field can be carefully balanced by an attractive potential due to a Kerr nonlinear medium, OFW solitons can be stably created.

Rydberg atoms with high polarizability and long-range interactions, have emerged as a robust platform to investigate optical soliton formation \cite{PRL.106.170401}. In this work, we propose a scheme for generating and stabilizing spatial OFW solitons in a cold Rydberg gas. The long-range {\it vdWs} interaction between Rydberg-state atoms induces giant nonlocal nonlinearity through electromagnetically induced transparency(EIT) effect which can compensate for the probe diffraction \cite{PRL.107.153001}. However this compensation is usually inadequate with an arbitrary interaction.
To achieve more stable OFW soliton we adopt the method of numerical optimization and identify the stability of such nonlocal OFW soliton can be greatly improved with both optimal atomic density and optimal one-photon detuning. A rough estimation shows that, the fidelity of OFW soliton preserves above $0.96$ for a propagation distance over 160 diffraction lengths.
Unlike previous variational approach that depends on a complex test solution to solve this problem \cite{OPTICA.6.000309}, our model provides a straightforward route to the production of spatial optical solitons with arbitrary intensity profiles.

\begin{figure}
    \centering
    \includegraphics[scale=0.4]{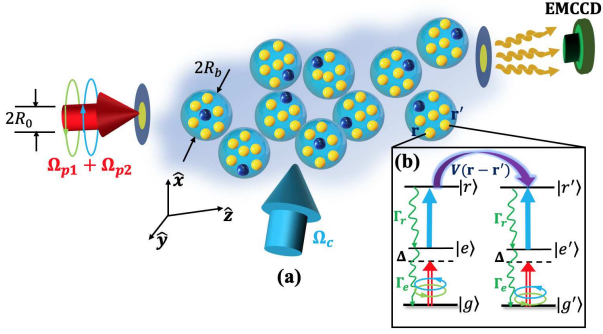}
    \caption{Propagation of an OFW field. (a) Schematic for the experimental demonstration. An atomic ensemble with a number of blocked Rydberg superatoms, is driven by a weak OFW field $\Omega_{p}(\bold{r})$ and a strong continuous field $\Omega_c$.
    $\hat{z}$ gives the propagation direction. The output intensity can be measured with an EMCCD. (b) Three-level atomic scheme. For $^{88}$Sr Rydberg atoms \cite{JPB:AMOP52(2019)244001(10pp)} with a density of $N_a=2\times10^{12}$ cm$^{-3}$, energy levels are  $|g\rangle=|5s^2~^1S_0\rangle$, $|e\rangle=|5s5p~^1P_1\rangle$ and $|r\rangle=|5s60s~^1S_0\rangle$. The spontaneous decay rate with respect to $|e\rangle$ is $\Gamma_e/2\pi = 16$ MHz; to $|r\rangle$ is  $\Gamma_r/2\pi = 16.7$ kHz. The coupling strengths are $\Omega_c/\Gamma_e=1.0$, $\Omega_{p0}/\Gamma_e=0.2$. }
    \label{fig1}
\end{figure}

{\it Model strategy-} We investigate the propagation stability of an OFW field that couples to a cold Rydberg atomic ensemble. 
Ensemble atoms are described by a three-level system shown in Fig.\ref{fig1}b in which the probe field with a Rabi frequency $\Omega_p(\bold{r})$(wavelength \textcolor{black}{461 nm}) couples the ground state $|g\rangle$ ($|g^{\prime}\rangle$) and the intermediate state $|e\rangle$ ($|e^{\prime}\rangle$); and the coupling field with $\Omega_c$(wavelength \textcolor{black}{420 nm}) couples $|e\rangle$ ($|e^{\prime}\rangle$) and the Rydberg state $|r\rangle$ ($|r^{\prime}\rangle$).
$\Delta$ is the intermediate-state detuning. Fig.\ref{fig1}a shows that, the probe field passing through the ensemble atoms has a collimated beam waist of $R_0$ while the strong coupling field is focused at the center of the atomic ensemble with a larger radius.
The structured pattern of the OFW field can induce a spatially dependent EIT where atoms are excited into Rydberg state at $\Omega_p(\bold{r})\neq 0$ \cite{PRL-114-123603}.
The probe field $\Omega_p(\bold{r})$ is weak and composed by two copropagating vortex beams 
\begin{equation}
  \Omega_{p}(\bold{r})=\Omega_{p1}(\bold{r})+\Omega_{p2}(\bold{r}),
  \label{1}
\end{equation} 
in which each field $\Omega_{pm}(\bold{r})$($m=1,2$) carrying OAM of $\hbar l_{m}$ per photon, is characterized by
\begin{equation}
  \Omega_{pm}(\bold{r}) =\Omega_{p0} \left(
  \bold{r}/R_0\right)^{|l_m|}e^{-\bold{r}^2/R_0^2}e^{il_m\phi}e^{ik_{p}{z}}. \label{2}
\end{equation}
Here $\bold{r}=(\bold{r_{\bot}},z)=(x,y,z)$, $\Omega_{p0}$ is a common laser amplitude, $\phi=\arctan(y/x)$ is the azimuthal angle and $k_{p}$ denotes the probe wave vector. $l_1=-l_2$ presents an OFW field.
The propagation distance is scaled by the diffraction length: $L_{diff}=k_{p}R_{0}^{2}$.

In the condition of rotating-wave approximation, the Hamiltonian of whole interaction system can be described by $\hat{H}=N_{a}\int\hat{H}_{s}(\bold{r}) d^{3}r$ where 
$\hat{H}_{s}= -\Delta\hat{\sigma}_{ee}-[\Omega_{p}(\bold{r})\hat{\sigma}_{ge}+\Omega_{c}\hat{\sigma}_{er}+h.c]
+N_{a}\{\int \hat{\sigma}_{r^\prime r^\prime}V(\bold{r}-\bold{r^{\prime}})
 d^{3}r^{\prime}\}\hat{\sigma}_{rr}$ 
represents the single-atom Hamiltonian.  $\hat{\sigma}_{\alpha\beta}=\vert\alpha\rangle\langle\beta\vert$ is the transition operator whose commutation relation obeys
$[\hat{\sigma}_{\alpha\beta},\hat{\sigma}_{u^\prime v^\prime}]=N_{a}^{-1}(\delta_{\alpha v'}\hat{\sigma}_{u'\beta}-\delta_{u' \beta}\hat{\sigma}_{\alpha v^\prime})\delta_{{rr^{\prime}}}$
($\delta_{rr^\prime}$
is the Kronecker symbol). $V(\bold{r}-\bold{r^{\prime}})=C_{6}/|\bold{r}-\bold{r}^{\prime}|^{6}$ denotes
the {\it vdWs} interaction between atoms at $\bold{r}$ and $\bold{r}^\prime$ and \textcolor{black}{ $C_6/2\pi=-81.6$ GHz$\cdot\mu$m$^6$ is the dispersion coefficient for state $|r^{(\prime)}\rangle$}. Without interaction ensemble atoms can be settled into a dark state such that the probe field $\Omega_p(\bold{r})$ is unaffected by the atomic medium \cite{PRL.107.153001}. However,
when the interaction is sufficiently strong it can inhibit multiple Rydberg excitations within a blockade radius of 
$R_{b}=(|C_{6}|/\delta_{\rm EIT})^{1/6}$ with $\delta_{\rm EIT}=\Omega_{c}^{2}/|\Delta+i\Gamma_e/2|$ the EIT linewidth, arising strong nonlinear optical modulation to the transmission of the probe OFW field \cite{PRL.107.133602}.

The time evolution of ensemble atoms can be governed by the master equation
\begin{equation}
\frac{\partial \hat{\rho}}{\partial t} =-\frac{i}{\hbar}\left[\hat{H},\hat{\rho}\right]+\Sigma_{j}\left[\Gamma_{j}\left(\hat{L}_{j}^\dagger\hat\rho \hat{L}_{j}-\frac{1}{2}\hat{L}_{j} \hat{L}^{\dagger}_{j}\hat\rho-\frac{1}{2}\hat{\rho} \hat{L}_{j} \hat{L}^{\dagger}_{j}\right)\right] \label{mas}
\end{equation}
where $\hat{\rho}=\sum_{u,v}|u\rangle\rho_{uv}\langle v|$ is the density matrix and $\rho_{uv}$ treats as the matrix elements describing the atomic
population for $u=v$ and coherence for $u\neq v$. Here $j=e,r$ and {$\hat{L}_e=\hat{\sigma}_{ge}$, $\hat{L}_r=\hat{\sigma}_{er}$}.
Upon the weak probe limit with $\Omega_{p0}\ll\Omega_c$
we employ a perturbation method expanded in $\Omega_{p0}/\Omega_c$ to solve Eq.(\ref{mas})(accurately up to $\Omega_{p}^{3}$). As a consequence one obtains an explicit expression for the steady state of matrix element  ${\rho}_{eg}$ that contains first-(linear relation) and third-order(nonlinear Kerr effect) solutions
\begin{eqnarray}
    \rho_{eg}(\bold{r})&\approx&\rho_{eg}^{(1)}\Omega_{p}(\bold{r})+\rho_{eg}^{(31)}|\Omega_{p}(\bold{r})|^{2}\Omega_{p}(\bold{r}) \nonumber  \\
    &+&N_{a}
    {\int}\rho_{eg}^{(32)}(\bold{r}_{\bot}-\bold{r_{\bot}^{\prime}})|\Omega_{p}(\bold{r_{\bot}^{\prime}},z)|^{2}d\bold{r}_{\bot}^{\prime} \Omega_{p}(\bold{r})
   \label{eg}
\end{eqnarray}
where we have assumed the spatial length of the probe field is sufficient large. A detailed expression of other matrix elements can be found elsewhere \cite{OE.24.004442}. 
After applying the multiple-scales method \cite{PRE.72.016617} all coefficients in Eq.(\ref{eg}) are analytically solvable,
\begin{eqnarray}
\rho^{(1)}_{eg}&=&\frac{2i\Gamma_{r}}{4\Omega_{c}^{2}+\Gamma^{\prime}\Gamma_{r}} \label{6} \\
\rho_{eg}^{(31)}&=&\frac{16i\Gamma_{r}^2[(4i\Omega_{c}^{2}+\Gamma_{r}(i\Gamma_{e}-2\Delta)]^{2}}{\Gamma_{e}(\Gamma_{e}^{2}\Gamma_{r}^{2}+8\Gamma_{e}^{2}\Omega_{c}^{2}+4\Gamma_{r}^{2}\Delta^{2}+16\Omega_{c}^{4})^{2}} \label{7} \\
\rho_{eg}^{(32)}(\bold{r_{\bot}}-\bold{r^{\prime}_{\bot}})&=&\frac{256(\Gamma^{\prime}+\Gamma_{r})\Omega_{c}^{4}}{(4i\Omega_{c}^{2}-\Gamma^{\prime}\Gamma_{r})|4\Omega_{c}^{2}+\Gamma^{\prime}\Gamma_{r}|^{2}} \nonumber
\end{eqnarray}
\begin{equation}
\times \int \frac{V(\bold{r}-\bold{r^{\prime}})d{z^{\prime}} }{4\Gamma^{\prime}\Omega_{c}^{2}+[\Gamma_{r}+i V(\bold{r}-\bold{r^{\prime}})][4\Omega_{c}^{2}-\Gamma^{\prime}(\Gamma^{\prime}+\Gamma_{r})]}\label{8}
\end{equation}
with $\Gamma^{\prime}=\Gamma_{e}-2i\Delta$. From Eq.(\ref{eg}) we know that the atomic coherence $\rho_{eg}(\bold{r})$ involves a nonlocal term $\propto \rho_{eg}^{(32)}(\bold{r}_\bot-\bold{r}_\bot^\prime)$ that depends on the strength of electric field at a more distant position $\bold{r}_\bot^\prime$ \cite{OE.425208}. While in a conventional EIT only a local nonlinear response works due to the photon-atom coupling rather than the Rydberg-Rydberg atom interaction \cite{PRL.107.213601}.

\begin{figure}
    \centering
    \includegraphics[width=8.8cm]{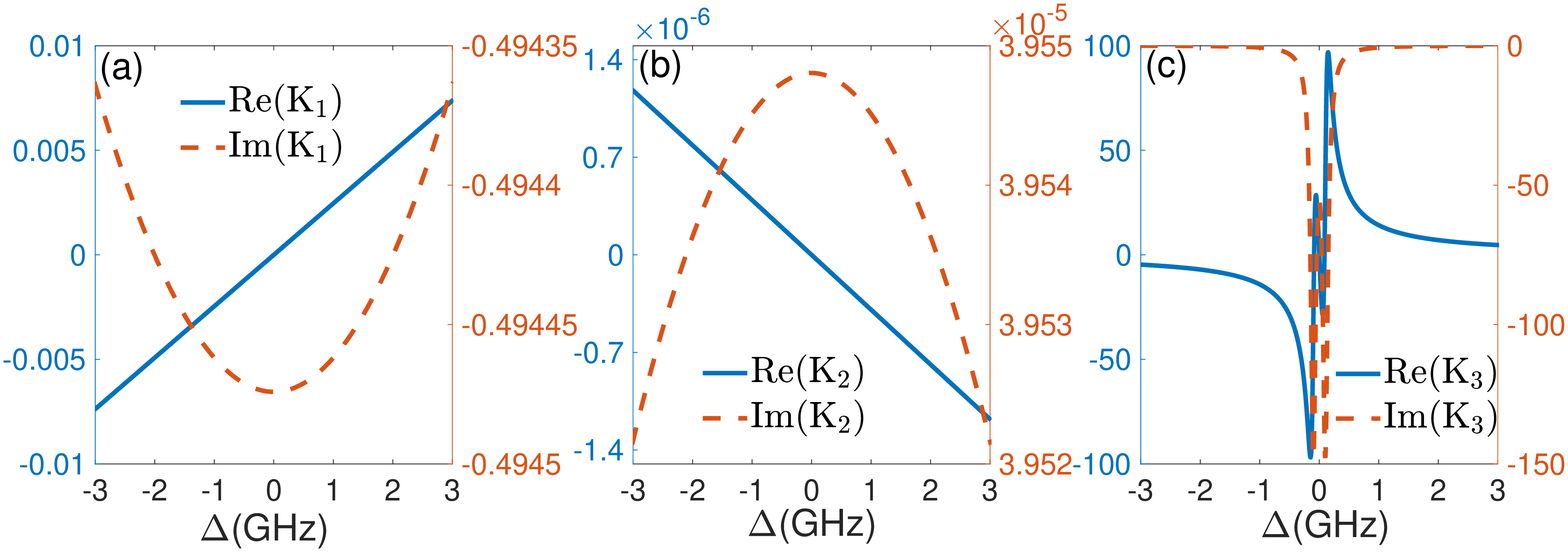}
    \caption{(a-c) Linear $K_1$, local nonlinear $K_2$, nonlocal nonlinear 
$K_3$ potentials {\it vs} the one-photon detuning $\Delta$. }
    \label{fig2}  
\end{figure}

{\it Different potentials-} Having established the basic theory for nonlocal Kerr nonlinearity of a strongly-interacting atomic ensemble, we now study the steady-state propagation of the probe field which is described by the Maxwell equation   
\begin{equation}
\left(\frac{\partial}{\partial z}-\frac{i}{2k_{p}}\nabla_{\bot}^{2} \right)\Omega_{p}(\bold{r})=i\kappa\rho_{eg}(\bold{r}) 
\end{equation}
with $\kappa=N_{a} k_{p}\mu^{2}/(\hbar\varepsilon_{0})$. Together with Eq.(\ref{eg}) one obtains a nonlinear Schr\"{o}dinger equation
\begin{eqnarray}
i{\partial_z U(\bold{r})}+{\nabla_{\bot}^{2}U(\bold{r})}/{2k_p}=-\kappa\rho^{(1)}_{eg}U(\bold{r})-\kappa \rho^{(31)}_{eg}|U(\bold{r})|^{2}U(\bold{r}) \nonumber \\
-\kappa N_{a}\int \rho^{(32)}_{eg}(\bold{r_{\bot}}-\bold{r_{\bot}^{\prime}})|U(\bold{r_{\bot}^{\prime}},z)|^{2} d \bold{r}_{\bot}^{\prime} U(\bold{r}) \label{11}
\end{eqnarray}
where 
\begin{equation}
U(\bold{r}_\bot,z=0)=\Omega_{p0}e^{-\bold{r_{\bot}}^2/R_0^2} \left[\left(
\bold{r}_{\bot}/R_0\right)^{|l_1|}e^{il_1\phi}+\left(\bold{r}_{\bot}/R_0\right)^{|l_2|}e^{il_2\phi}\right] \nonumber
\end{equation}
is the incident field.
In Eq.(\ref{11}), the transverse Laplacian $\nabla_{\bot}^{2}$ term contributed by the kinetic energy of ensemble atoms leads to probe diffraction. Terms related to $\rho_{eg}^{(1)}$, $\rho_{eg}^{(31)}$, $\rho_{eg}^{(32)}$ respectively, account for the linear, local nonlinear and nonlocal nonlinear potentials. Especially the nonlocal potential depending on $N_a^2$ can be made greatly attractive(due to $C_6<0$) for a large atomic density, which is crucial for the formation of stable OFW solitons.

Figure \ref{fig2} compares the magnitudes of three different potentials as a function of the detuning $\Delta$:
\begin{equation}
 K_1 = -{\kappa\rho_{eg}^{(1)}}/{\Omega_{p0}^2  },K_2=-\kappa\rho_{eg}^{(31)},K_3 = -\kappa N_a\int\rho_{eg}^{(32)}({R})d{R}
 \nonumber
\end{equation}
with ${R}=\bold{r}_{\bot}-\bold{r}_{\bot}^\prime$ the relative distance along radial direction. Due to the strong interaction, we observe that, in the range of $\Delta\in[-3,3]$ GHz the nonlocal nonlinear potential $|K_3|$ is greatly larger than linear $|K_1|$ and local $|K_2|$ potentials by orders of magnitude, and thus the terms $\propto K_1,K_2$ are negligible. Therefore Eq.(\ref{11}) is reduced to
\begin{equation}
i\frac{\partial U(\bold{r}) }{\partial z}+\frac{\nabla_{\bot}^{2}U(\bold{r})}{2k_{p}}
+\kappa N_{a}\int \rho^{(32)}_{eg}(R)|U(\bold{r_{\bot}^{\prime}},z)|^{2} d \bold{r}_{\bot}^{\prime} U(\bold{r})=0 \label{re}
\end{equation}
Note that when the probe diffraction can be carefully balanced by a self-focusing attractive potential $Re(K_3)<0$ ($Im(K_3)$ is suppressed because of a large one-photon detuning), stable light soliton can be formed. From Fig.\ref{fig2}c we know it can be achieved by properly selecting a negative $\Delta$.

\begin{figure}
    \centering
    \includegraphics[scale=0.245]{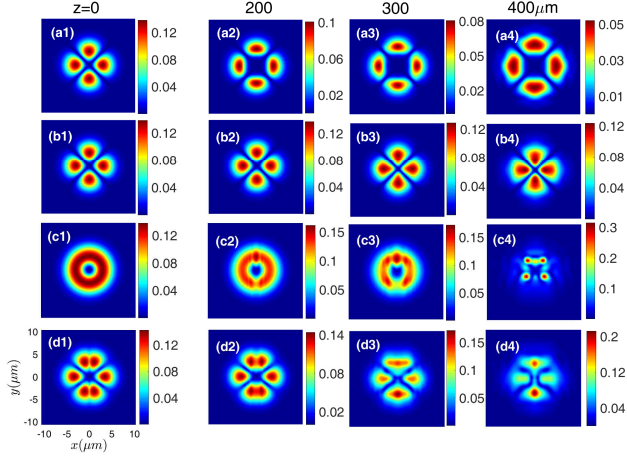}
    \caption{Radial
 intensity profiles $|U(\bold{r}_\bot,z)|^2/\Omega_{p0}^2$
 at different propagation distances $z=(0, 200,300,400)$ $\mu$m. From top to bottom, we use $(l_1,l_2)=(2,-2)$, $(2,-2)$ for an OFW field and $(l_1,l_2)=(2,2)$, $(2,-3)$ for an unmatched composite vortex field, and $\Delta=(2,-2,-2,-2)$ GHz, accordingly.}
    \label{fig3}  
\end{figure}

{\it Propagation of the probe field-} We now turn to investigate the nonlinear propagation of an OFW field. To ensure that the probe field can feel the modulation performed by a fairly strong nonlinear potential we choose 
$R_0=3$ $\mu$m to fit $R_0<R_b$, leading to $L_{diff}\approx122.665$ $\mu$m.
By evolving Eq.(\ref{re}) the output intensity profiles under different pairs of winding numbers $(l_1,l_2)$ and detunings $\Delta$, are displayed in Fig.\ref{fig3}. 
In the case of OFW fields with equal opposite $l_1=-l_2=2$($l=|l_{1,2}|$),
panels (a1-a4) and (b1-b4) show the output intensity profiles under different signs of detunings. $\Delta = (2,-2)$ GHz individually present a repulsive and attractive nonlocal potential.
 In general we observe that, the probe-field profile can be greatly preserved without distinct deformation in these two cases. Because optical phase information has been converted into intensity information by interfering two vortice in producing OFW \cite{OE.15.008619}. 
The probe intensity has a symmetrically azimuthal dependence 
$2 \cos(2\phi)$ and the imaginary phase part in the input field $U(\bold{r}_{\bot},z=0)$ has been annihilated.
However, as $z$ increases, the shape of probe field reveals a clear expansion accompanied by a big reduction in the peak intensity [see (a4)], which is mainly caused by the nature of nonlocal interaction. When $\Delta=2$ GHz the combination of a repulsive nonlocal potential and the probe diffraction leads to strong radial spread.
As turning to $\Delta=-2$ GHz in (b1-b4), an attractive nonlocal potential with $Re(K_3)<0$ can partially balance the probe diffraction arising more stable propagation. Nevertheless, (b4) shows that the intensity profile has a small convergence, because $\Delta$ is arbitrarily chosen here. To ensure stable long-distance propagation of the OFW field, a careful optimization of parameters is necessary.

For comparison we also explore the case with equal $l_1=l_2$ and non-equal $|l_1|\neq|l_2|$. It is clear that the output intensities can not keep their shape due to a large nonlocal potential. At $z=400$ $\mu$m it reveals a catastrophic collapse accompanied by the localization of intensities at certain points [see (c4) and (d4)]. This is because such composite fields still contain singularity splitting 
as a pure vortex beam which leads to the singularities distributed around the space
\cite{OE.20.022961}. 

\begin{figure}
    \centering
    \includegraphics[scale=0.28]{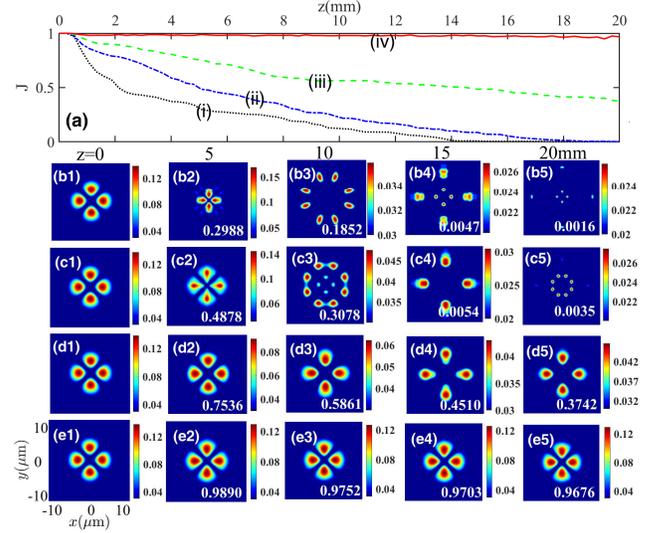}
    \caption{Long-distance propagation of the OFW field($l=2$) under optimization. (a) Stability factor $J$ {\it vs.} the longitudinal distance $z\in[0,20]$ mm(equivalent to $[0,163]L_{diff}$).
    Cases (i-iv) described by the dotted, dashed-dotted, dashed and solid lines, correspond to $\Delta=(-2.0,-2.8403,-2.0,-2.19617)$ GHz, $N_a = (2.0,2.0,1.3802,1.4748)\times 10^{12}$ cm$^{-3}$. Accordingly, (b-e) show the time evolution of the radial intensity profiles at $z=(0,5,10,15,20)$ mm. $J$ values are given in the lower right of figures.}
    \label{fig:my_label}  
\end{figure}

{\it Long-distance propagation by parameter optimization-}
 We employ the genetic algorithm to optimize parameters in the nonlocal potential $K_3$ allowing for a straightforward balance between the probe diffraction and the nonlocal nonlinear potential. 
To quantify the propagation stability we introduce an overlap integral $J$ serving as the stability criterion
\cite{PhysRevA.78.021802}
\begin{equation}
J=\frac{|\int U(\bold{r_{\bot}},{z})U(\bold{r_{\bot}},{z}=0)d\bold{r}_{\bot}|^{2}}{\int|U(\bold{r_{\bot}},{z})|^{2}d\bold{r}_{\bot} \int |U(\bold{r_{\bot}},{z}=0)|^{2}d\bold{r}_{\bot}} \label{12}
\end{equation}
where $J\to 1$ means a high-fidelity propagation without deformation. The principle of genetic algorithm is illustrated elsewhere \cite{PRA.17.024014}. Except $\Delta$, we treat $N_a$ as a second optimization parameter due to $K_3\propto N_a^2$.
After performing sufficient runs of optimization 
we show the variation of $J$ as well as the time evolution of the probe fields in Figure~\ref{fig:my_label}(a) and (b-e).
 It is evident that, in cases (i)(non-optimization) and (ii)(single-optimization of $\Delta$) which correspond to panels (b1-b5) and (c1-c5), the shape of the probe field changes significantly exhibiting an entire splitting effect. When $z>10$ mm the OFW field is localized at several points where the average intensity at each localization suffers from a big reduction. At $z=20$ mm, the propagation fidelity $J$ is much smaller than 0.01 with a peak intensity around $0.026$. It is remarkably that, adding a suitable adjustment of $N_a$[see (d1-d5)] has a profound influence on the probe propagation.
The pattern of the OFW field can be well preserved against collapse, in spite of a small spread caused by the imperfect balance between the attractive nonlocal potential and the probe diffraction.

Most importantly, once both $\Delta$ and $N_a$ are optimized at the same time we see the stability factor $J$(red-solid) in (a) can sustain at a high level $>0.96$. This is verified in panels (e1-e5) where the spatial evolution of the probe intensity profile is very stable which means we have realized stable OFW soliton via parameter optimization in a nonlocal Rydberg medium. The way of parameter optimization can serve as a robust tool for achieving stable OFW solitons against expansion or convergence.

 \begin{figure}
    \centering
    \includegraphics[scale=0.3]{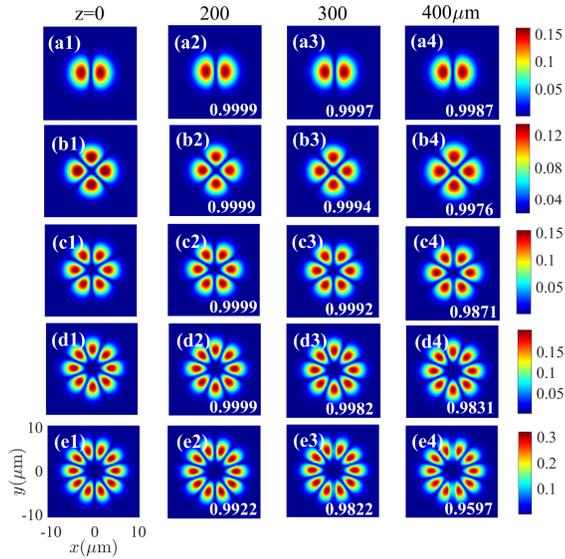}
    \caption{Stable propagation of higher-order OFW lights with different winding numbers. From top to bottom $l=(1,2,3,4,5
)$, $\Delta = (-2.523$, $-2.9617$, $-1.113$, $-0.5832$, $-0.5545)$ GHz, $N_a=(0.8261$, $1.4748$, $0.5693$, $0.3415$, $0.2851$)$\times 10^{12}$ cm$^{-3}$. }
    \label{fig5}  
\end{figure}

{\it Stable propagation of higher-order OFW fields -} The OFW created by higher-order vortex beams behaves as petal modes which provides more flexible ways for atomic localization. In Fig. \ref{fig5} we study the propagation of radial intensity profiles of various OFW fields. With optimal parameters the final intensity profiles can preserve their petal shapes with a high fidelity.
While as $l$ increases we see a small decrease in $J$ occurs because the OFW having a larger beam radius can not be sufficiently modulated by a finite atomic nonlocal potential. 
If the probe-field radius is extremely large the nonlocal effect contributed by the Rydberg-Rydberg interaction will reduce into a local Kerr nonlinearity \cite{OPTICA.6.000309}.
Luckily, accounting for the fact that this potential also depends on the strength of probe and coupling lasers we find an auxiliary optimization to both $\Omega_{p0}$ and $\Omega_c$ could significantly enhance the propagation stability of higher-order OFW fields. A rough estimation implies,
when $\Omega_c/\Gamma_e = 0.7579$, $\Omega_{p0}/\Gamma_e = 0.1356$ and keep optimal $\Delta$, $N_a$ values, the $J$ factor for $l=5$ at $z=400$ $\mu$m($\approx 3.3L_{diff}$) can be increased to 0.9896, which means the way of multi-target optimization could further improve the stability of higher-order OFW solitons.

{\it Conclusions-}We have studied the formation of spatial OFW solitons when a weak probe OFW field propagates through a cold Rydberg gas. By the virtue of Rydberg EIT effect the strong {\it vdWs} interaction between Rydberg-state atoms can induce a giant nonlocal potential which can compensate for the diffraction of the OFW field. However this compensation is usually inadequate leading to a short propagation distance. Via an optimal adjustment for both the atomic density and the one-photon detuning we are able to achieve
stable OFW soliton with a propagation distance over 160 diffraction length and the stability preserving above 0.96.
Formation of arbitrary higher-order OFW solitons is also possible by using a robust optimization taking account of more parameters.
A pure vortex beam is known to carry OAM and suffers from singularity splitting during its propagation. While we note that, the incoherent superposition of two OFWs can counter-intuitively produce a vortex beam \cite{OL.39.000704}.
So our results may offer an alternative way for realizing stable propagation of pure optical vortices in a strong nonlocal atomic medium.

~\\
\textbf{Funding.} National Natural Science Foundation of China (12174106,
11474094, 12247146).

~\\
\textbf{Disclosures.} The authors declare no conflicts of interest.

~\\
\textbf{Data availability.} Data underlying the results presented in this paper are
not publicly available at this time but may be obtained from the authors upon
reasonable request.


\begin{thebibliography}{52}
\makeatletter


\bibitem{OE.20.022961}
F.~Ricci, W.~L{\"o}ffler, and M.~Van~Exter, {\protect\JournalTitle{Optics
  Express}} \textbf{20}, 22961 (2012).

\bibitem{NJP.20-043023}
M.~P. Lavery, {\protect\JournalTitle{New Journal of Physics}} \textbf{20},
  043023 (2018).

\bibitem{PRL.129.073902}
G.~W. Henderson, G.~R. Robb, G.-L. Oppo, and A.~M. Yao,
  {\protect\JournalTitle{Physical Review Letters}} \textbf{129}, 073902 (2022).

\bibitem{OE.22.009920}
R.~Neo, S.~J. Tan, X.~Zambrana-Puyalto, S.~Leon-Saval, J.~Bland-Hawthorn, and
  G.~Molina-Terriza, {\protect\JournalTitle{Optics Express}} \textbf{22}, 9920
  (2014).

\bibitem{OPTICA.3.000355}
A.~Hansen, J.~T. Schultz, and N.~P. Bigelow, {\protect\JournalTitle{Optica}}
  \textbf{3}, 355 (2016).

\bibitem{PRA.96.013622}
V.~E. Lembessis, {\protect\JournalTitle{Physical Review A}} \textbf{96}, 013622
  (2017).

\bibitem{OE.15.008619}
S.~Franke-Arnold, J.~Leach, M.~J. Padgett, V.~E. Lembessis, D.~Ellinas, A.~J.
  Wright, J.~M. Girkin, P.~{\"O}hberg, and A.~S. Arnold,
  {\protect\JournalTitle{Optics Express}} \textbf{15}, 8619 (2007).

\bibitem{JOSAB.443903}
V.~Lembessis, A.~Lyras, and O.~Aldossary, {\protect\JournalTitle{JOSA B}}
  \textbf{38}, 3794 (2021).

\bibitem{OL.427000}
H.~R. Hamedi, V.~Kudria{\v{s}}ov, N.~Jia, J.~Qian, and G.~Juzeli{\=u}nas,
  {\protect\JournalTitle{Optics Letters}} \textbf{46}, 4204 (2021).

\bibitem{nanoph-2021-0814}
B.~Mao, Y.~Liu, W.~Chang, L.~Chen, M.~Feng, H.~Guo, J.~He, and Z.~Wang,
  {\protect\JournalTitle{Nanophotonics}} \textbf{11}, 1413 (2022).

\bibitem{PRL.106.170401}
F.~Maucher, N.~Henkel, M.~Saffman, W.~Kr{\'o}likowski, S.~Skupin, and T.~Pohl,
  {\protect\JournalTitle{Physical Review Letters}} \textbf{106}, 170401 (2011).

\bibitem{PRL.107.153001}
S.~Sevin{\c{c}}li, N.~Henkel, C.~Ates, and T.~Pohl,
  {\protect\JournalTitle{Physical Review Letters}} \textbf{107}, 153001 (2011).

\bibitem{OPTICA.6.000309}
Z.~Bai, W.~Li, and G.~Huang, {\protect\JournalTitle{Optica}} \textbf{6}, 309
  (2019).

\bibitem{PRL-114-123603}
N.~Radwell, T.~W. Clark, B.~Piccirillo, S.~M. Barnett, and S.~Franke-Arnold,
  {\protect\JournalTitle{Physical Review Letters}} \textbf{114}, 123603 (2015).

\bibitem{PRL.107.133602}
A.~V. Gorshkov, J.~Otterbach, M.~Fleischhauer, T.~Pohl, and M.~D. Lukin,
  {\protect\JournalTitle{Physical Review Letters}} \textbf{107}, 133602 (2011).

\bibitem{JPB:AMOP52(2019)244001(10pp)}
F.~Robicheaux, {\protect\JournalTitle{Journal of Physics B: Atomic, Molecular
  and Optical Physics}} \textbf{52}, 244001 (2019).

\bibitem{OE.24.004442}
Z.~Bai and G.~Huang, {\protect\JournalTitle{Optics Express}} \textbf{24}, 4442
  (2016).

\bibitem{PRE.72.016617}
G.~Huang, L.~Deng, and M.~Payne, {\protect\JournalTitle{Physical Review E}}
  \textbf{72}, 016617 (2005).

\bibitem{OE.425208}
Y.-L. Zhou, {\protect\JournalTitle{Optics Express}} \textbf{29}, 15300 (2021).

\bibitem{PRL.107.213601}
D.~Petrosyan, J.~Otterbach, and M.~Fleischhauer,
  {\protect\JournalTitle{Physical Review Letters}} \textbf{107}, 213601 (2011).

\bibitem{PhysRevA.78.021802}
I.~Novikova, N.~B. Phillips, and A.~V. Gorshkov,
  {\protect\JournalTitle{Physical Review A}} \textbf{78}, 021802 (2008).

\bibitem{PRA.17.024014}
R.~Li, S.~Li, D.~Yu, J.~Qian, and W.~Zhang, {\protect\JournalTitle{Physical
  Review Applied}} \textbf{17}, 024014 (2022).

\bibitem{OL.39.000704}
I.~A. Litvin, S.~Ngcobo, D.~Naidoo, K.~Ait-Ameur, and A.~Forbes,
  {\protect\JournalTitle{Optics Letters}} \textbf{39}, 704 (2014).

\end{thebibliography}

\end{document}